# Valuation methods for professional sports clubs: A historical review, a model development, and the application to Japanese football clubs

Masaaki Kimura[a], Zen Walsh[b], Takuo Inoue[a]✉, Toshiya Takahashi[b], Hideki Koizumi[b]

[a] Research Center for Advanced Science and Technology, The University of Tokyo;
[b] Graduate School of Engineering, The University of Tokyo

✉Corresponding Author: Takuo Inoue (Email: tak@up.t.u-tokyo.ac.jp )

## Abstract

**Research question**: In the trend towards the globalization of football and the increasing commercialization of professional football clubs, a methodology for calculating the firm value of clubs in non-western countries has yet to be established. This study reviews the valuation methods for the club firm values in Europe and North America and how values are calculated at the time of changing ownership of Japanese clubs and develops regression models with high explanatory power to estimate the more accurate firm value of Japanese football clubs.

**Research methods**: A review of the existing literature on methods for calculating the firm value of professional sports clubs in Europe and North America, as well as financial statements and registers relating to changes of ownership of Japanese clubs, was conducted. After that, multiple regression analyses were conducted using the KPMG's enterprise value of European clubs as the explained variable.

**Results and Findings**: From the literature review and the Japanese case studies, it has become clear that the standard valuation methods of European clubs are based on revenue, plus taking into account factors such as stadium ownership, wage ratio, operating profit, net assets, player market value, among other variables. In contrast, in Japan, valuation is based solely on the par value of stocks or net assets. Multiple regression analysis revealed that the firm value of European clubs over the past three years is best explained by revenue or player market value and the number of SNS followers.

**Implications**: Two models with high explanatory power were developed, the estimated firm value using the revenue-based formula being higher than that based on player market value. However, in the J.League, the former was more than three times higher than the latter, while the former was only 1.2 times higher for European clubs. It was suggested that the discrepancy relates to differences in European and J.League clubs' revenues and asset structures. In either formula, the firm value of J.League clubs exceeded the actual transaction price when the change of ownership occurred in the past.

## Chapter 1: Introduction

It is indisputable that football is now the world's most popular sport. According to the world governing body FIFA, the 2022 FIFA World Cup saw nearly 1.5 billion people worldwide watch the final (FIFA, 2022). The U.S.-based business magazine Forbes reported that the event had been expected to generate USD 4.7 billion in revenue for FIFA (Forbes, 2022). The fact that 211 countries and territories initially entered the tournament's qualifying round and 206 teams played the matches indicates that the sport is played worldwide.

The globalization of football is manifested not merely as the distribution of the playing population to all regions but also as economic globalization through the advent of multinational major clubs (Giulianotti & Robertson, 2004). Football clubs that generate huge profits attract the attention of investors, and ownership changes are occurring actively around the world, leading to the commercialization of football clubs (Dubal, 2010; Rohde & Breuer, 2017; Wilson et al., 2013). It cannot be denied at this time that professional sports have become a huge industry and have significantly impacted the economies of the cities and states in which they are based. Geopolitical changes must also be addressed to capture the tide of commercialized football clubs (Chadwick, 2022), considering owners in countries such as Saudi Arabia, Qatar, the UAE, and China are attempting to change the dynamics of professional football from a European focus.

Under these circumstances, even outside Europe, the crux of the football club market, there has been an increasing number of changes in the ownership of football clubs. In 2018, when an ownership change occurred at Kashima Antlers in Japan's J.League, it was suggested that the price was 'too low' for this deal, bringing about the widespread discussion in the Japanese football community about whether the club's firm value had been appropriately evaluated. Without a proper firm valuation calculation method, purchasing and selling professional sports clubs would be conducted irrationally. In addition, it is also difficult to raise necessary funding foundations, and consequently, the market cannot grow aggressively. This situation creates a lost opportunity for current club owners, leaving Japan behind in global trends. Calculating the appropriate firm value of clubs in emerging football countries outside the highly priced European club market is vital for sound global football.

Several things could be addressed in assessing the value of non-European professional football clubs like Kashima Antlers of Japan and confirming an intuitive judgment that they are 'too low.' First, it needs to be clarified how the firm values of clubs have been calculated when buying and selling Japanese professional football clubs. It is also necessary to clarify the methods used globally in calculating the value of clubs and their applicability to Japan, which is a prerequisite for discussing the appropriate method of calculating the value of professional football clubs in Japan. Market research is also necessary, as the price will ultimately be determined by investors and owner companies who intend to buy. As part of the efforts to develop appropriate valuation methods applicable to professional football clubs in non-European countries, including Japan as an example, this paper aims to review existing valuation methods, confirm the current method of valuing clubs when they are bought or sold in Japan, and formulate models for calculating the firm value of professional football clubs with greater explanatory power to estimate the actual firm value of Japanese clubs. In addition, by applying the proposed valuation model, the firm value of each J.League club is estimated, and the characteristics of Japanese clubs are discussed compared to those of their European counterparts.

The study is structured as follows. First, Chapter 2 provides a literature review of the valuation methods by which the firm value of football clubs has been estimated globally. The chapter will first summarize the methods used to assess firm value, which is not limited to football clubs, and then describe the discussion in determining which methods have been deemed suitable for football clubs and the specific models employed. The differences in methods for calculating firm values in the two primary forms of professional sports league structures, European and North American, are also examined. In Chapter 3, while considering the result of the literature review of firm value calculation methods, the current state of value calculation in Japan is clarified by inferring from publicly available information how those values were calculated in the actual case of management control changes in Japan. Based on the structure and situation of Japanese clubs and leagues, Chapter 4 insists that adopting the European rather than the North

American model as a comparator for Japan's valuation method is appropriate. In Chapter 5, based on previous studies and publicly available information, the variables believed to be included in existing value calculation models are listed to identify the variables and their coefficients with the highest explanatory power through statistical analysis. In Chapter 6, the proposed value calculation models obtained in Chapter 5 are applied to Japanese clubs to calculate the estimated firm value of each club. Finally, Chapter 7 discusses Japanese clubs' characteristics and the significance of the model development.

## Chapter 2: A historical review of valuation methods

*Commercialization of football clubs in Europe*

How are the firm values of clubs estimated globally, and what specific valuation methods are used when there are control changes? Before conducting a historical review on this point, it is necessary to consider the trend toward commercializing professional football clubs. One of the significant events that led to the accelerated commercialization of clubs was the establishment of the English Premier League (EPL) in 1992 and the consequent commencement of the active acceptance of club ownership by foreign companies and private owners.

In Europe, until the 1990s, many people held club shares out of a sense of psychological satisfaction, contribution to the community, or a sense of duty to the community rather than for economic benefits. It made the private (joy of victory and honor of being supported) and social (duty as a personage) satisfaction factors significant; therefore, owners were sometimes called 'trophy owners' (Geckil et al., 2007; Tiscini & Dello Strologo, 2016). Vine (2004) and Geckil et al. (2007) refer to factors other than economic benefits as 'ego factors' and insist they are hard to quantify. Huth (2020) also indicates that attachment to the club was essential for investment in football clubs and argues that investment in football clubs was partly independent of economic benefits, which is still the case today.

However, the situation behind club ownership has drastically changed since the establishment of EPL. The new owners are no longer tied to the region and are unlikely to hold the club permanently. It was essential to increase the economic value of the club given future management transfers, which inevitably led to the need to assess the firm value of football clubs as accurately as possible and to increase that value. Accordingly, discussions on how to define and capitalize the previously ambiguous player market values have progressed (Morrow, 1999; Amir & Livne, 2005), and since 2005, the German data company Transfermarkt has been publishing the value of each player belonging to European clubs on an annual basis. At the same time, Forbes began publishing the World's Most Valuable Soccer Teams, judged by the estimated firm values, annually since 2004 (Forbes, 2024). In the meantime, to deter bankruptcies caused by profligate management, leagues introduced financial licensing systems in the early 2010s, leading to improved financial stability. All these not only brought stability and transparency to the management of European football clubs and attracted new investors from all over the world but also promoted the development of clubs' firm valuation methods and accelerated the trend of management transfers by clubs that had increased their value (Nauright & Ramfjord, 2013; Thani & Heenan, 2017). The recognition of the need for an accurate firm valuation of football clubs, the subject of this research, has advanced considerably in Europe over the past two decades.

### *Commonly used valuation methods and their applicability to football clubs*

This section provides an overview of the methods commonly used to calculate the firm value of companies, regardless of the enterprise domain, plus diving deep into their applicability to football clubs. Though various approaches and their classification frameworks have been proposed, they comprise income, market, and cost approaches (Aydin, 2017; Damodaran, 2024).

In the income approach (not always but commonly Discounted Cash Flow Method), one of the most typical approaches, the expected profit for the next three to five years is discounted from the present value. In Europe, no football league has a salary cap system with which clubs are made to comply in any country (Markham, 2013), meaning that player labor costs do not have a ceiling. In these circumstances, the inverse correlation between competitive performance and player labor costs increases, making it difficult to win without increasing player labor costs, and good competitive performance does not lead to profit (Hamil & Walters, 2010; Storm & Nielsen, 2012). The relationship between performance and profit is a trade-off, with 'profit maximization' being the emphasis on profit without being willing to lower performance and 'win maximization' being the emphasis on performance without being willing to lose money (Garcia-del-Barrio & Szymanski, 2009; Leach & Szumanski, 2015; Storm & Nielsen, 2012). Then, managers would want to let owners avoid making a profit, as in most cases, the desire to avoid relegation outweighs. Hamil & Walters (2010) splendidly showed that in the 17 years of EPL, from its inception in 1992 until they published their paper, there was not a single year when the club's total income and expenditure were in surplus. This unique circumstance makes it difficult to adopt income approaches to evaluation based on profit (Deloitte, 2012; cited in Markham, 2013).

Another widely accepted category of valuation methods is the market approach (sometimes called relative valuation). However, there are several reasons why the market approach is not considered appropriate for calculating the firm value of football clubs. First, Markham (2013) reported that there were only 22 listed clubs in Europe at the time of writing his paper, and even in 2024, with some changes in numbers, the number is still minor at approximately 15 clubs. It makes a sample size too small to use the market approach, a relative valuation method calculated by reference to the value of companies in the same industry. Secondly, listed shares are often purchased and held by fans and supporters, who will not sell their shares once they have held them, so those who try to take control have to pay a substantial premium to the stock price if they want to acquire the number of shares needed (La Porta et al., 1997). Club shares are, therefore, illiquid and do not reflect balance sheet figures. The small sample size and low liquidity due to share ownership by fans make it difficult to calculate the value using the market approach and sometimes lead to trading at prices that are considered to be significantly higher than the actual value. For example, when Tottenham was delisted in 2012, its market capitalization was 830 thousand euros, while the valuation based on trading volume was 2.45 million euros, and the valuation by Forbes was 3.51 million euros, causing a significant difference in estimates due to differences in methodology. In addition, the actual trading value was reportedly higher than the Forbes price at the time (Geckil et al., 2007; Markham, 2013). As mentioned above, even though the stock price is generally an important indicator when measuring the value of a company, it is not appropriate to use this calculation method as the football club's stock price does not represent a fair value.

One other method of valuation is the cost approach (also known as intrinsic valuation). Although net assets are a critical value assessment factor (Sanchez et al., 2022), no articles were found explaining their use in calculating the firm value of football clubs. Their inability to reflect the club's profitability, brand, and potential has hindered incorporating this line of methods as the primary ones of club valuation.

### *Focus on revenue in valuing football clubs*

This section reviews the methods used to calculate football clubs' firm value. First, many papers commonly emphasize the importance of revenue rather than the aggregate of future profits (Geckil et al., 2007; Harris, 2006; Markham, 2013; Sanchez et al., 2022;

Scelles et al., 2013; Scelles et al., 2016; Tiscini & Dello Strologo, 2016). Some have stated that valuation methods based on revenue are optimal and have even developed specific calculation formulae. The expression of firm value as a revenue multiple is also used in business cases (Damodaran, 2012) based on the fact that the increase in firm value (capital gains) usually results from a situation where sales growth exceeds accumulated losses and losses carried forward (income losses). There was an old and controversial case of this kind of calculation. David Dein, who became the owner of Arsenal, bought the club for much more than the former owner, Peter Hill-Wood, bought it. At that time, Dein was derided as using 'dead money' as he seemed to buy the club at a cost that was too high. However, Dein transferred his shares at 2.56 times higher than the acquisition price afterward (Moore, 2006; Sanchez et al., 2022). The accountant firm Deloitte, positioned to comprehend the actual transaction price, also places importance on revenue and other sales-related figures, reporting that the transaction price typically costs the equivalent of 1.5 and 2.0 times the annual revenue (Deloitte, 2012; cited in Markham, 2013).

In a U.S. case study, Fort (2006) points out that for the past 40 years, the growth rate of team sales prices has been more than twice the economic growth. The idea that 'sales growth, rather than income and expenditure, has the greatest impact on company value,' assuming steady sales growth, has become the standard valuation method for promising tech start-ups and is also noted as the best valuation method for football clubs at present (Markham, 2013; Sanchez et al., 2022; Scelles et al., 2016). In all three papers, the firm value of clubs was used as the explained variable. Multiple regression analysis investigates which factors strongly influence, and revenue was always included as a variable.

### *Variables included in the European models*

This section offers a more detailed look at the existing valuing methods in Europe. A review of the literature on valuation methods in European football shows that variables other than revenue differ in the literature. Markham (2013) uses revenue, net assets, net profit, attendance, and team wage cost ratio as variables and derives his calculation formula. Tiscini et al. (2016) use revenue, Earnings Before Interest and Taxes (EBIT), Return On Assets (ROA), Enterprise Value to Earnings Before Interest and Taxes ratio (EV/EBIT), and Enterprise Value to Sales multiple (EV/Sales) as variables when conducting multiple regression analysis. While acknowledging the impact of revenue, the results pointed out that these variables did not sufficiently explain firm values, so psychological factors (ego factors) were inevitably added. Sanchez et al. (2022), also aware of firm valuation in the U.S. as described in the next section, conducted a multiple regression analysis using hinterland population, stadium age, broadcasting rights fees, net assets, debt equity ratio, domestic GDP, and domestic and international competition results as variables. As a result, they argued that revenue and net assets are essential factors. KPMG, a leading chartered accountancy firm, also publishes the Football Benchmark report, which annually reports the estimated firm value of major European clubs. They determine firm values based on six factors: revenue, player market value, the number of followers on social networking services (SNS), broadcasting rights revenue, wage cost ratio, and stadium ownership. They say their firm value estimation is based on actual transaction prices, as they can access many examples.

### *Comparison with North American models: similarities and differences*

Turning to the U.S. and Canada, where other major professional sports are taking place, there is also literature on the firm valuation of the clubs. While there are a small number of listed clubs in Europe, as referred to above, there are only a few in the four major professional sports leagues in North America (American football, basketball, baseball, and ice hockey), which considerably limits the financial data publicly available. Therefore, development of valuation models has to rely only on the empirical sale price of the teams (Humphreys & Lee, 2010; Humphreys & Mondello, 2008), and it is difficult to find literature presenting the calculation method. However, even in this context, similarities and differences with Europe can be found.

Similarities include the fact that the rate of increase in firm value exceeds the economic growth rate, the capital gain in firm value also exceeds the team's accumulated deficit, revenue-based valuation methods dominate actual transactions, and the existence of ego factors (c.f. Vine, 2004; Fort, 2006; Humphreys & Mondello, 2008).

A notable difference is whether 'profit' is considered an explanatory variable. Regarding league orientation, North American leagues aim to 'maximize league profits.' In contrast, Europe intends to 'optimize the profits of individual clubs' (Leach & Szymanski, 2015). This difference is manifested in the following features that exist only in North American leagues and not in European football: there is a player salary cap (NFL, 2024), a draft system, and no promotion or relegation. In addition, the NFL even has a revenue-sharing system. With these league structures, North American leagues are intended to balance the competitiveness between teams and maximize overall profits at the expense of the maximum profit potential of some leading individual teams (Alexander & Kern, 2004; Fort & Quirk, 1995; Leach & Szymanski, 2015; Nauright & Ramfjord, 2010). It has been argued that operating profits are one factor in assessing firm value, as teams are operated under profit-oriented management (Scelles et al., 2016; Tistini & Dello Strologo, 2016). Some papers also argue that franchise population is one factor influencing the calculation of firm value because the U.S. domestic market is closed and does not allow multiple franchises to be established in the same metropolitan area, in addition to the lack of intra- and inter-continental competition (Alexander & Kern, 2004; Humphreys & Mondello, 2008).

Thus, the standard view in Europe and North America is to evaluate revenue as the primary explanatory variable. In the meantime, due to the different structures of the leagues, some differences were found in the explanatory variables other than revenue.

## Chapter 3: Previous cases on the valuation of Japanese football clubs

So, what kind of club valuation methods have been used in the transactions of club management rights in Japan? This chapter meticulously investigates the transaction prices of Japanese professional football clubs at the time of change in ownership. Over the eight years from 2017 to 2024, 18 ownership changes occurred in 17 clubs in the J.League. The acquisition of control is achieved through the purchase of shares, with two primary patterns. The first is the outstanding shares transfer, and the second is the issue and acquisition of new shares. Of the 18 cases mentioned above of ownership changes over the past eight years, the relevant figures are listed for all cases for which information is available from published figures, registers, and official gazettes. In particular, the list includes the acquisition pattern, the amount of money required for the purchase, par value, stock price, the amount required to purchase 51% of all shares, and the method of deciding the transaction price (**Table 1**).

**Table 1**. Critical aspects of the acquisition of control during 2017-2024 in J.League

| Club | Pattern of Acquisition | Par Value (k¥) | Stock Price (k¥) | Amount of Money Required to Purchase 51% of the Share (m¥) | Assumed Method of Transaction Price |
|---|---|---|---|---|---|
| FC Tokyo | Capital increase | 50 | 50 | 1,200 | Par value |
| FC Machida Zelvia | Capital increase | 50 | 50 | 714 | Par value |
| Sagan Tosu | Share transfer | 10 | 3 | NA | Net asset |
| Kashima Antlers | Share transfer | 50 | 82.7 | 1,330 | Net asset-based |

When Japanese IT company Mixi, Inc. acquired the management rights of FC Tokyo (the operating company's name is Tokyo Football Club) in 2021, the acquisition method was a capital increase. The stock price was 50 thousand yen per share, and management rights were obtained by purchasing 24,000 shares, or 51.3% of the total shares. FC Tokyo's capital situation at the end of FY 2018 and FY 2021, as published by the J.League, shows a capital of 1.187 billion yen and a capital surplus of 0. According to the register, the number of shares was 23,740 in 2018, and this has not changed in 2021, so the price per share before the capital increase is 1.187 billion yen / 23,740 shares = 50 thousand yen. The capital and number of shares in October 1998, when the company was founded, are not recorded in the register. However, the data published by J.League shows that the earliest figure was 783 million yen in 2005, then 807 million yen in 2006, 815 million yen in 2007, 1.005 billion yen in 2010, and 1.065 billion yen in 2013. With so many capital increases, it is unlikely that the stock price has remained a round number when calculated at anything other than par value. Therefore, it is likely that the stocks have consistently traded at 50 thousand yen per share since the club's foundation. In 2018, the Japanese advertising company CyberAgent Inc.'s approach to acquiring control of FC Machida Zelvia was also a capital increase, with 1.148 billion yen invested at 50 thousand yen per share. It has been made clear that 82% of the voting rights were acquired, indicating that 714 million yen was needed to acquire 51%, which would have enabled the acquisition of management rights. As in the case of FC Tokyo, the transaction was carried out at a par value of the shares.

On the other hand, the approach adopted when Best Amenity Co, Ltd. acquired a 47.3% stake in Sagan Tosu (the operating corporation Sagan Dreams) in January 2021 was to transfer shares from the previous largest shareholder. The amount of the transfer was not disclosed, but the register confirms that Sagan Dreams increased its capital by 150,180 shares in July 2021 following the transfer; the data J.League published for FY2020 and FY2021 shows an increase in capital and capital reserves of 4,505.4 thousand yen, so it is assumed that the capital increase was done at the rate of 3,000 yen per share (4,505,400/150,180). The register for January 2020 shows that the number of shares was 84,500, the capital was 218.975 million yen, and the capital surplus at that time was 203.9 million yen from the J.League data publicly available, so 425,064/84,500 = 5,000 yen per share, indicating that the shares were traded at a lower price than their par value at the time of the share transfer. Although it is hard to know how the capital increase was calculated, it can be presumed it was not based on profits but on net assets, as the company was close to insolvency. The case of Kashima Antlers, the one mentioned in Chapter 1, also involved a transfer of management rights through a share transfer. On this occasion, 61.6% of all shares were acquired for 1.6 billion yen. The number of shares available from the register is 31,400, which means 82,700 yen per share; the J.League data shows that 61.6% of net assets at the time of the transaction was 1.334 billion yen, so the actual transaction price was close to this. On the other hand, the net profit for FY 2018 was 425 million yen, which makes it difficult to believe that the valuation method based on profit was used. As the transaction was carried out at a higher price than the par value calculated using the net asset method, it is possible that some factors were taken into account based on net assets.

As the investigation has shown, there are two primary acquisition approaches: those traded at a fixed price (par value) of shares and those based on net assets. As they are based on par value or net assets, the amount of money needed to gain management rights is influenced by the total number of shares issued up to the point of acquisition. These shreds of evidence show that in the selling and purchasing of Japanese football clubs, there were no attempts to calculate the firm value of clubs more proximate to reality by combining various variables based on revenue, as in the European and North American examples explained in Chapter 2.

## Chapter 4: Context of J.League: parallels to European leagues

Existing literature has shown that valuation methods for football clubs and other professional sports teams are becoming established in Europe and North America. This chapter considers whether these methods can be applied to Japanese clubs. It is necessary to consider whether European or North American leagues and Japan's J.League can be said to be in the same industry and whether applying the same firm valuation method to them is appropriate.

### *Product*

First of all, the leagues mentioned above are the same in that they deal with the sport of football. Matches last 90 minutes and are played between 11 players each, including a goalkeeper. The leagues comprise tens of clubs ranked in a yearly league competition. The product is football entertainment, for which talented players are acquired.

### *Promotion/relegation systems*

One of the main elements that make Japan's J.League different from North America and similar to Europe is the existence of a promotion/relegation system. Firm value largely varies depending on whether a team belongs to the top category, which is directly related to revenue and the acquisition and retention of players. In addition, as discussed in Chapter 2, the existence of promotion and relegation creates a tendency to make it harder to make a profit, which prevents the introduction of the income approach as a method of valuation. Aiming for increased revenues and an efficient wage ratio becomes more critical, as revenue is directly related to winning.

### *Salary cap*

As discussed in Chapter 2, European football has no salary cap. The same is true in Japan, where it is not stated in the league regulations (J.League, 2024c). On the other hand, Maximum Salary Budget Charge, an equivalent of a salary cap, has been introduced in Major League Soccer in the U.S. (MLS, 2024). Introducing a luxury tax would likely impact firm valuations as it discourages acquiring highly paid players.

### *Revenue structure*

As seen in **Table 2**, the primary sources of revenue generated from league operations are broadcasting rights, matchday, and sponsorship, both in European leagues and J.League (Deloitte, 2023; J.League, 2024a). Matchday revenues mainly consist of ticket revenues, VIP room sales, and merchandise revenues. Note that the J.League's 'Distribution Fee' is paid mainly from broadcasting rights fees, and 'other commercial' is integrated under 'Sponsorship/Commercial' in England, Spain, and Italy. The identicality of primary revenue and expenditure items is recognized between Europe and Japan. Aside from the amounts, the fact that the components of revenue and expenditure are almost equal is essential for considering valuation methods.

**Table 2**. Revenue structure of European leagues and J.League clubs on average

| Item | England | Spain | Germany | Italy | France | Japan |
|---|---|---|---|---|---|---|
| Matchday | 14% | 12% | 9% | 10% | 11% | 18.9% |
| Broadcasting | 54% | 59% | 44% | 57% | 36% | - |
| Distribution Fee | - | - | - | - | - | 6.5% |
| Sponsorship/Commercial | 32% | 29% | 29% | 33% | 32% | 41.3% |
| Other commercial | - | - | 18% | - | 21% | 33.2% |

### *Financial licensing*

To discourage financial profligacy, club license systems are introduced, under which licenses cannot be issued if the club has been in the deficit for three consecutive years or if the accumulated deficit for three years exceeds a certain amount unless a certain level of net assets exists. More specifically, the financial criteria in Europe only allow the issuance of the license if 'the cumulative deficit over three years does not exceed €5 million, or if the club can recapitalize in the year and the cumulative deficit does not exceed €60 million' ("UEFA Club Licensing and Financial Sustainability Regulations" Article 82.02; UEFA, 2023). A license cannot be issued in Japan if the club has been 'in the deficit for three consecutive years and is insolvent' ("J.League Financial Standards" F.01 3; J.League, 2024b). There are some subtle differences, but they are the same: A license cannot be issued if the club has a deficit in its accumulated income and expenditure over three years.

Considering all these factors, there are sufficient reasons why the same framework for firm valuations can be applied to European football clubs and J.League clubs.

## Chapter 5: Development of highly agreeing firm value models

In this chapter, formulae for calculating the firm value of football clubs are derived by multiple regression analysis, using the estimated firm value of European football clubs as the explained variable, as well as financial figures and other data of each club as explanatory variables.

**Tables 3** and **4** list the analyzed clubs and the indicators used as the candidates of explanatory variables. In this study, the enterprise value (EV) published by KPMG was used as the explained variable, which contains most of the explanatory variables adopted in the previous studies described in Chapter 2.

**Table 3**. List of European football clubs analyzed in this paper

| Country | Club | Country | Club |
|---|---|---|---|
| England | Arsenal | Netherlands | Ajax |
| England | Aston Villa | France | Olympique Lyonnais |
| England | Chelsea | France | Olympique Marseille |
| England | Everton | France | Paris Saint-Germain |
| England | Leicester City | Germany | Bayern Munich |
| England | Liverpool | Germany | Borussia Dortmund |
| England | Manchester City | Germany | Eintracht Frankfurt |
| England | Manchester United | Germany | Schalke 04 |
| England | Tottenham Hotspur | Italy | AC Milan |
| England | West Ham United | Italy | AS Roma |
| Spain | Athletic Bilbao | Italy | Atalanta |
| Spain | Atletico Madrid | Italy | Inter Milan |
| Spain | Barcelona | Italy | Juventus |
| Spain | Real Madrid | Italy | SS Lazio |
| Spain | Sevilla | Italy | SSC Napoli |
| Spain | Valencia | Türkiye | Besiktas |
| Spain | Villareal | Türkiye | Fenerbahce |
| Portugal | Benfica | Türkiye | Galatasaray |
| Portugal | Porto | | |

**Table 4**. List of the candidates as explanatory variables

| Item | Data source |
|---|---|
| Total followers on major Social Networking Services | Includes: X (formerly Twitter), Instagram, Facebook, YouTube, Weibo, TikTok (As of 25 January 2024) |
| Revenue | KPMG (2021) |
| Player market value | Transfermarkt, 2023-24season (As of January 2024) |
| Broadcasting rights revenue | KPMG (2021) |
| Wage Cost Ratio | KPMG (2021) |
| Player wages | Capology (As of 26 January 2024) |

Multiple regression analysis was conducted on these data. After increasing and decreasing explanatory variables to derive regression equations with high explanatory power, two formulae with high correlation coefficients were derived using a combination of significant explanatory variables (**Table 5**). SNS followers and revenue were adopted as explanatory variables in Formula 1, while SNS followers and player market value in Formula 2.

**Table 5**. Two formulae developed in this study

(i) SNS & Revenue Model

| Statistics | | | Coeff. | Standard Error | t Stat | P-value |
|---|---|---|---|---|---|---|
| Multiple R | 0.9878 | Intercept | 0 | | | |
| R Square | 0.9758 | SNS Followers (m) | 3.7233 | 0.6486 | 5.7410 | 1.69.E-06 |
| Adjusted R Square | 0.9466 | Revenue (m€) | 2.9233 | 0.2284 | 12.8016 | 9.17.E-15 |
| Standard Error | 223.7454 | | | | | |

(ii) SNS & Player Market Value Model

| Statistics | | | Coeff. | Standard Error | t Stat | P-value |
|---|---|---|---|---|---|---|
| Multiple R | 0.9749 | Intercept | 0 | | | |
| R Square | 0.9505 | SNS Followers (m) | 5.7754 | 0.7994 | 7.2246 | 1.96.E-08 |
| Adjusted R Square | 0.9205 | Player Market Value (m€) | 1.2599 | 0.1599 | 7.8815 | 2.89.E-09 |
| Standard Error | 320.1838 | | | | | |

**Table 6** shows the estimated firm values of big European football clubs calculated using these formulae. Overall, the EV published by KPMG and the average firm value estimated using the developed formula that includes player market value (FV2) were almost identical. In contrast, results from the formula that includes revenue (FV1) were around 5% higher. Although some clubs showed a considerable deviation in valuations between different calculation methods, similar values were calculated by both regression formulae, especially for clubs with a high EV.

**Table 6**. Estimated firm values of European football clubs using the formulae developed (unit: m€)

| Club | EV (KPMG) | FV1 | FV2 | Club | EV (KPMG) | FV1 | FV2 |
|---|---|---|---|---|---|---|---|
| Real Madrid | 3,184 | 3,283 | 3,500 | Ajax | 473 | 461 | 395 |
| Manchester United | 2,883 | 2,469 | 2,251 | Lyon | 456 | 437 | 299 |
| Barcelona | 2,814 | 3,072 | 3,215 | Atalanta | 454 | 498 | 437 |
| Bayern | 2,749 | 2,290 | 1,971 | Everton | 450 | 687 | 507 |
| Liverpool | 2,556 | 2,139 | 1,928 | Eintracht Frankfurt | 428 | 495 | 314 |
| Manchester City | 2,483 | 2,431 | 2,476 | Roma | 413 | 675 | 596 |
| Chelsea | 2,179 | 2,003 | 2,139 | Sevilla | 390 | 544 | 291 |
| PSG | 2,132 | 2,399 | 2,464 | Valencia | 385 | 359 | 312 |
| Tottenham | 1,912 | 1,525 | 1,518 | Besiktas | 383 | 286 | 250 |
| Juventus | 1,597 | 1,837 | 1,411 | Galatasaray | 344 | 355 | 533 |
| Arsenal | 1,584 | 1,454 | 1,994 | Athletic Bilbao | 336 | 338 | 359 |
| Atletico Madrid | 1,234 | 1,195 | 828 | Benfica | 326 | 311 | 536 |
| Dortmund | 1,226 | 1,198 | 888 | Porto | 311 | 484 | 395 |
| Inter Milan | 996 | 1,241 | 1,094 | Aston Villa | 308 | 653 | 887 |
| AC Milan | 578 | 925 | 1,065 | Villareal | 303 | 400 | 277 |
| West Ham | 541 | 698 | 647 | Lazio | 302 | 491 | 327 |
| Leicester | 526 | 841 | 430 | Marseille | 195 | 499 | 437 |
| Schalke | 502 | 469 | 74 | Fenerbahce | 184 | 337 | 436 |
| Napoli | 483 | 571 | 744 | | | | |

## Chapter 6: Adoption of the developed models to J.League clubs

The two formulae introduced in Chapter 5 were then used to calculate the firm value of J.League clubs. The results are shown in **Table 7**. For clubs mentioned in Chapter 3, calculation results by Formula 1 were 304-603% higher than actual transaction prices, while the results by Formula 2 were 65-77% higher, when the exchange rate was assumed to be 150 yen to the euro. Urawa Reds, which had the highest revenue and player market value of all J.League clubs, was the highest valued club by both formulae. Its firm value based on Formula 1 is around half of major Turkish clubs, and approximately 5-10% of Europe's highest valued clubs. For all clubs, the estimated firm value using Formula 1 was significantly higher than the value based on Formula 2, with the revenue-oriented firm value being more than three times higher on average. **Figure 1** indicates a clear difference in trends of evaluation results depending on the calculation method between European clubs and J.League clubs. J.League clubs also have a lower correlation between the two firm values compared to the European clubs included in this research.

**Table 7**. Firm values of J.League clubs based on regression models

| League (2023) | Club | SNS followers | Revenue (m€) | Player market value (m€) | FV1 (m€) | FV2 (m€) | FV1/FV2 |
|---|---|---|---|---|---|---|---|
| J1 | Hokkaido Consadole Sapporo | 412,622 | 24.03 | 13.98 | 71.79 | 20.00 | 359.0% |
| J1 | Kashima Antlers | 792,968 | 40.77 | 20.80 | 122.14 | 30.79 | 396.8% |
| J1 | Urawa Reds | 807,734 | 54.18 | 28.55 | 161.39 | 40.64 | 397.2% |
| J1 | Kashiwa Reysol | 205,307 | 30.88 | 12.85 | 91.03 | 17.38 | 523.9% |
| J1 | FC Tokyo | 664,305 | 35.16 | 17.50 | 105.26 | 25.89 | 406.6% |
| J1 | Kawasaki Frontale | 1,276,055 | 46.53 | 22.18 | 140.76 | 35.31 | 398.6% |
| J1 | Yokohama F. Marinos | 858,356 | 43.21 | 18.65 | 129.50 | 28.45 | 455.1% |
| J1 | Yokohama FC | 208,682 | 19.07 | 11.64 | 56.53 | 15.87 | 356.2% |
| J1 | Shonan Bellmare | 311,333 | 16.51 | 15.03 | 49.43 | 20.73 | 238.4% |
| J1 | Albirex Niigata | 312,910 | 16.93 | 9.50 | 50.65 | 13.78 | 367.6% |
| J1 | Nagoya Grampus | 748,573 | 40.61 | 17.12 | 121.49 | 25.89 | 469.2% |
| J1 | Kyoto Sanga F.C. | 171,441 | 21.92 | 16.33 | 64.72 | 21.56 | 300.1% |
| J1 | Gamba Osaka | 583,844 | 39.79 | 16.95 | 118.50 | 24.73 | 479.2% |
| J1 | Cerezo Osaka | 1,575,578 | 28.11 | 18.79 | 88.03 | 32.77 | 268.6% |
| J1 | Vissel Kobe | 742,724 | 42.43 | 27.18 | 126.81 | 38.53 | 329.1% |
| J1 | Sanfrecce Hiroshima | 530,245 | 26.78 | 17.78 | 80.26 | 25.46 | 315.2% |
| J1 | Avispa Fukuoka | 223,352 | 18.86 | 10.95 | 55.96 | 15.09 | 371.0% |
| J1 | Sagan Tosu | 267,045 | 18.41 | 8.85 | 54.80 | 12.69 | 431.8% |
| J2 | Vegalta Sendai | 195,377 | 17.77 | 11.65 | 52.68 | 15.81 | 333.3% |
| J2 | Blaublitz Akita | 51,435 | 5.85 | 5.18 | 17.28 | 6.82 | 253.3% |
| J2 | Montedio Yamagata | 160,809 | 14.61 | 9.55 | 43.32 | 12.96 | 334.2% |
| J2 | Iwaki FC | 87,485 | 5.13 | 0.65 | 15.33 | 1.32 | 1157.8% |
| J2 | Mito Hollyhock | 131,046 | 6.83 | 3.95 | 20.44 | 5.73 | 356.6% |
| J2 | Tochigi SC | 105,261 | 6.94 | 6.28 | 20.68 | 8.52 | 242.7% |
| J2 | Thespa Gunma | 64,961 | 4.78 | 4.58 | 14.22 | 6.15 | 231.3% |
| J2 | Omiya Ardija | 157,217 | 17.59 | 10.88 | 52.00 | 14.62 | 355.8% |
| J2 | JEF United Ichihara Chiba | 170,435 | 17.59 | 7.48 | 52.05 | 10.41 | 500.0% |
| J2 | Tokyo Verdy | 630,280 | 14.11 | 6.98 | 43.58 | 12.43 | 350.5% |
| J2 | FC Machida Zelvia | 93,469 | 12.79 | 12.00 | 37.75 | 15.66 | 241.1% |
| J2 | Ventforet Kofu | 122,516 | 10.43 | 9.18 | 30.94 | 12.27 | 252.1% |
| J2 | Zweigen Kanazawa | 100,941 | 5.75 | 6.68 | 17.19 | 9.00 | 191.1% |
| J2 | Shimizu S-Pulse | 310,083 | 33.91 | 18.80 | 100.29 | 25.48 | 393.7% |
| J2 | Jubilo Iwata | 270,856 | 21.55 | 11.80 | 64.00 | 16.43 | 389.5% |
| J2 | Fujieda MYFC | 59,559 | 2.70 | 3.24 | 8.11 | 4.43 | 183.3% |
| J2 | Fagiano Okayama | 116,314 | 12.55 | 6.33 | 37.11 | 8.65 | 429.2% |
| J2 | Renofa Yamaguchi FC | 93,151 | 7.45 | 7.05 | 22.13 | 9.42 | 235.0% |
| J2 | Tokushima Vortis | 112,151 | 14.81 | 11.65 | 43.72 | 15.33 | 285.3% |
| J2 | V-Varen Nagasaki | 168,653 | 13.76 | 16.37 | 40.85 | 21.60 | 189.1% |
| J2 | Roasso Kumamoto | 101,617 | 6.52 | 2.58 | 19.44 | 3.84 | 506.5% |
| J2 | Oita Trinita | 161,032 | 12.18 | 8.64 | 36.20 | 11.82 | 306.4% |
| J3 | Vanraure Hachinohe | 29,921 | 2.67 | 2.05 | 7.93 | 2.76 | 287.6% |
| J3 | Iwate Grulla Morioka | 44,261 | 4.48 | 4.29 | 13.26 | 5.66 | 234.3% |
| J3 | Fukushima United | 30,776 | 2.87 | 2.39 | 8.51 | 3.19 | 267.0% |
| J3 | Y.S.C.C. Yokohama | 25,568 | 1.05 | 1.88 | 3.17 | 2.52 | 126.2% |
| J3 | S.C. Sagamihara | 137,554 | 5.08 | 2.28 | 15.36 | 3.67 | 418.9% |
| J3 | Matsumoto Yamaga F.C. | 212,459 | 10.07 | 5.40 | 30.22 | 8.03 | 376.3% |
| J3 | AC Nagano Parceiro | 50,775 | 5.05 | 3.42 | 14.96 | 4.60 | 325.1% |
| J3 | Kataller Toyama | 53,703 | 4.51 | 3.55 | 13.39 | 4.78 | 280.0% |
| J3 | Azul Claro Numazu | 42,229 | 2.89 | 1.86 | 8.62 | 2.59 | 333.0% |
| J3 | FC Gifu | 104,264 | 5.85 | 4.30 | 17.48 | 6.02 | 290.4% |
| J3 | FC Osaka | 32,822 | 3.73 | 1.96 | 11.02 | 2.66 | 414.3% |
| J3 | Nara Club | 35,188 | 2.86 | 1.73 | 8.49 | 2.38 | 356.4% |
| J3 | Gainare Tottori | 67,383 | 3.24 | 2.92 | 9.72 | 4.07 | 239.0% |
| J3 | Kamatamare Sanuki | 59,808 | 2.71 | 2.17 | 8.14 | 3.08 | 264.2% |
| J3 | Ehime FC | 60,826 | 5.25 | 5.58 | 15.58 | 7.38 | 211.1% |
| J3 | FC Imabari | 57,486 | 6.97 | 6.27 | 20.58 | 8.23 | 250.0% |
| J3 | Giravanz Kitakyushu | 71,613 | 6.82 | 4.35 | 20.20 | 5.89 | 342.8% |
| J3 | Tegevajaro Miyazaki | 23,804 | 2.17 | 2.32 | 6.42 | 3.06 | 209.8% |
| J3 | Kagoshima United FC | 77,314 | 5.06 | 3.77 | 15.08 | 5.20 | 290.2% |
| J3 | FC Ryukyu | 79,501 | 10.66 | 5.55 | 31.46 | 7.45 | 422.2% |
| | Average | 257,583 | 15.4 | 9.2 | 46.0 | 13.1 | 342.0% |
| | Median | 126,781 | 11.4 | 7.0 | 33.8 | 9.9 | 333.1% |

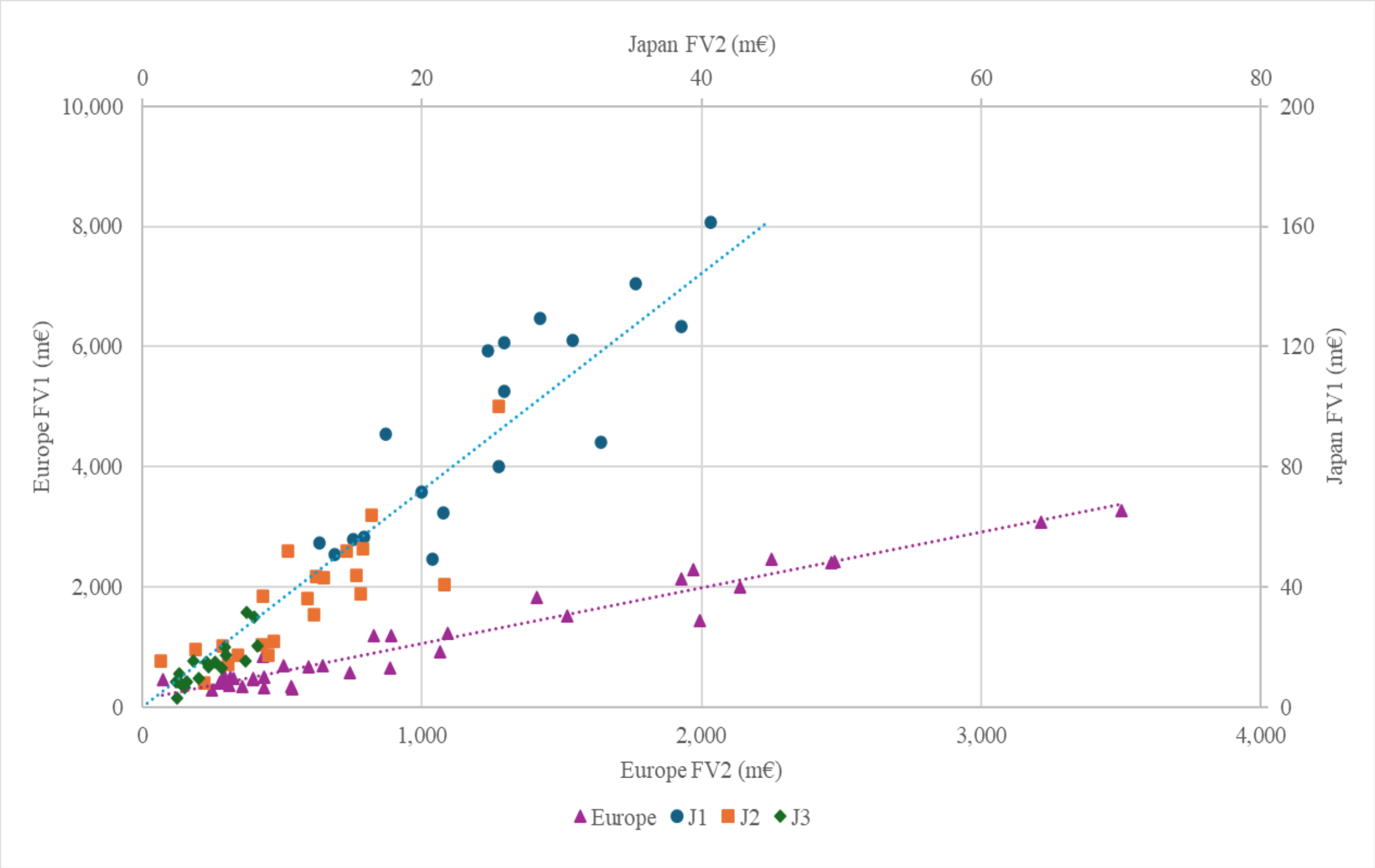

**Figure 1**. Firm values of J.League and European clubs

**Figure 2** shows the relationship between calculation results from the two formulae. Clubs such as Kashiwa Reysol, Gamba Osaka and Iwaki FC have higher FV1 than FV2. On the other hand, Cerezo Osaka, which has the most SNS followers, and V-Varen Nagasaki are among those who have a high FV2 in comparison with FV1. In general, the two formulae show relatively close valuations for clubs in lower categories.

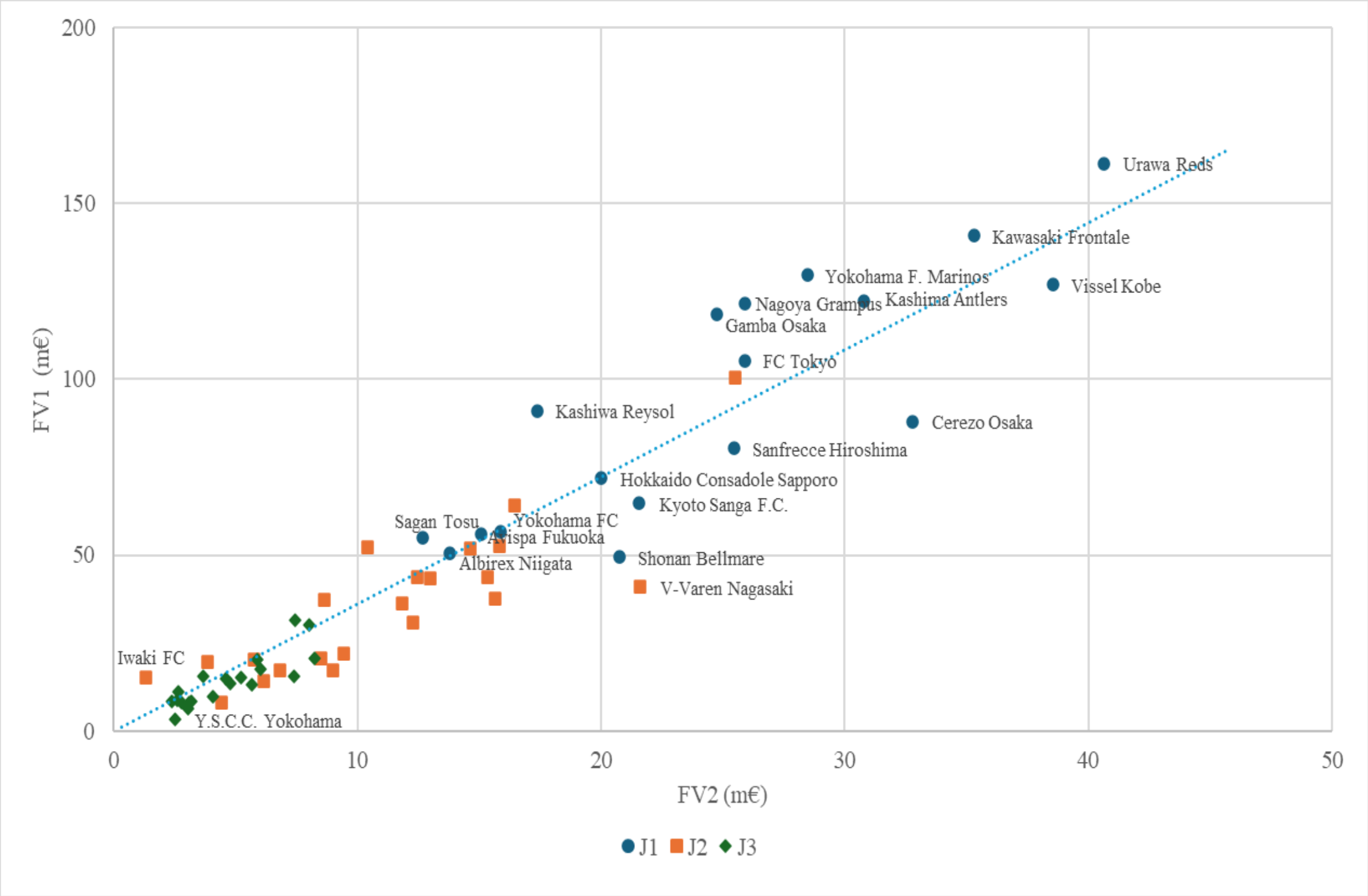

**Figure 2**. Relationship between calculated firm values of J.League clubs

## Chapter 7: Comparative analysis of football club valuation methods

In this paper, intending to develop an appropriate valuation method applicable to professional football clubs in non-European countries such as Japan, existing valuation methods were reviewed first. Next, the methods of deciding transactional prices used in Japan when buying and selling clubs was investigated. Models for calculating the firm value of professional football clubs with greater explanatory power were then developed, and these proposed formulae were applied to estimate the accountable firm value of each J.League club. Based on these results, this chapter discusses the characteristics of Japanese clubs compared to European clubs after summarizing what has been made clear.

Chapters 2 to 4 investigate the methods used to value professional sports clubs in Japan, Europe, and North America. A historical review of European and North American case studies found that in Europe and North America, a standard method of calculating firm values have included factors such as stadium ownership, team wage ratio, operating profit, net assets, and player market value, among others, with a shared focus on revenue. On the other hand, it was considered that J.League clubs had experienced changes of ownership with the valuations based on the par value of shares or the net assets, which could be confirmed using the data published by J.League and the corporate registry. In addition, a comparison of the structure of European leagues and J.League was made to conclude that it would be reasonable to apply the European valuation methods to the Japanese situation.

Based on the above, models for calculating the firm value of professional football clubs with higher explanatory power were developed in Chapter 5. Formulae were created using data from the last three years about the factors discussed in previous studies as explanatory variables, and KPMG's EVs, derived from knowledge of actual transaction prices, as the explained variable. As a result, two highly explanatory formulae were developed: revenue and the number of SNS followers, and player market value and the number of SNS followers. The number of SNS followers had more influence on the valuation as a factor than revenue. The fact that both formulae included SNS followers suggests that the existing method of valuations based on revenue, described in much previous research, may be beginning to change for big clubs. The importance of the number of SNS followers has been remarked on in the past (Scelles et al., 2013). With many football leagues having started to sell their broadcasting rights outside their own country, the number of fans worldwide is also an impactful factor in shaping club value. In European clubs, which now boast a global profile, their popularity and brand value cannot be explained by attendance numbers alone, and the argument that the number of SNS followers has a significant impact on firm value is in line with recent trends.

The estimated firm values of European and J.League clubs, calculated using the two developed formulae, revealed the following characteristics of European and Japanese clubs. For the J.League clubs, FV1 was more than three times higher than FV2 on average across all clubs, while FV1 was 1.2 times higher than FV2 on average when the two formulae were applied to the European clubs. This discrepancy is thought to be related to the difference in the amount of revenue and asset structures between European clubs and J.League clubs: it can be imagined that J.League clubs generate higher sales compared to European clubs that have similar player market value and SNS followers, but this also suggests the need for J.League clubs to increase their players' market values in the European market in order to increase firm value in the future.

Differences in valuations per calculation formula were also observed, indicating the characteristics of the clubs: in Chapter 6, it was observed that clubs located in the top left of the graph in **Figure 2** were characterized by higher revenue than is expected considering the market value of the players they own. On the other hand, clubs located in the bottom right of the graph were characterized by higher player market value for the revenue obtained. It may suggest differences in strategies between clubs and could be a topic for future research.

This paper successfully developed firm valuation formulae based on the existing European framework to apply to estimate the firm value of Japanese professional football clubs. Deriving the calculation formulae based on examples from Europe and applying it to Japanese clubs to estimate their values were meaningful in making the foundation for the discussion of the appropriate valuation of the clubs. However, it remains an issue that while the European valuation models are based on actual transactional prices, it is yet to be known whether some people or companies would buy Japanese clubs at the prices estimated using the formulae proposed in this paper. It is hoped that future research will be conducted to clarify this.

## Disclosure statement

This work was funded by Meiji Yasuda Life Insurance Company.